\documentclass[aps,amsmath,amssymb,twocolumn,10pt,prc]{revtex4-1}
\usepackage[T1]{fontenc}
\usepackage{graphicx}  
\usepackage[colorlinks=true,linkcolor=blue,citecolor=red,urlcolor=magenta]{hyperref}

\begin{document}  
  
\title{Fundamental limitations of the eigenvalue continuation approach}
\author{Tomasz Sowi\'nski$^{1,2}$ and Miguel A. Garcia-March$^{2}$} 
\affiliation{\mbox{$^1$Institute of Physics, Polish Academy of Sciences, Aleja Lotnik\'ow 32/46, PL-02668 Warsaw, Poland}
\mbox{$^2$Instituto Universitario de Matem\'atica Pura y Aplicada, Universitat Polit\`ecnica de Val\`encia, ES-46022 Val\`encia, Spain}}
\date{\today} 
 
\begin{abstract}
In this work, we show that the eigenvalue continuation approach introduced recently in [Phys. Rev. Lett. {\bf 121}, 032501 (2018)], despite its many advantages, has some fundamental limitations which cannot be overcome when strongly correlated many-body systems are considered. Taking as a working example a very simple system of several fermionic particles confined in a harmonic trap we show that the eigenvector continuation is not able to go beyond the accuracy of the sampling states. We support this observation within a very simple three-level model capturing directly this obstacle. Since mentioned inaccuracy cannot be determined self-consistently within the eigenvalue continuation approach, support from other complementary methods is needed. 
\end{abstract}  
\maketitle  

\section{Introduction}
One of the most challenging tasks in describing strongly correlated many-body systems relies on the accurate determination of their many-body spectra. Even if one considers very simplified theoretical models describing interacting particles, finding an appropriate numerical approach to capture their eigenstates is very hard. The main obstacle comes from the fact that the size of the corresponding Hilbert space grows exponentially with the number of particles considered, while interactions couple all (or almost all) of them \cite{1971FetterBook}. Thus, the many-body basis of any approximate method has to be adequately tailored to capture the appropriate Hilbert subspace containing desired many-body state. It was recently shown that a very useful method for this purpose is the eigenvalue continuation method. In short, the method originates on the observation that for smoothly varying Hamiltonians their temporal eigenstates are well-captured in the basis spanned by eigenstates obtained for other values of varying parameters~\cite{2018FramePRL}. This method has been used successfully in a variety of systems~\cite{2020DemolPRC,2021DrischlerPhysLettB,2020KonigPhysLettB}, and recent studies have focused on the determination of its numerical convergence~\cite{2021SarkarPRL}.

In this work, our aim is to shed some light on the limitations of this widely explored method. Our motivation is based on the observation that the fundamental reasoning standing behind eigenvalue continuation, {\it i.e.}, smoothness of states on control parameters, is in practice not always sufficient to obtain well-converged results. The reason is that the final accuracy (accuracy of the eigenstate obtained for the extrapolated parameter) depends substantially on the accuracy of the sampling states. Even if the location in the Hilbert space of the sampling states is well-determined it does not necessarily mean that the target state obtained by smooth extrapolation is well characterized by their linear combination. With two exemplary models discussed in the following we show that, along with changing control parameters, the target state may quickly flow out from the subspace spanned by sampling states and lead to inaccurate or simply wrong results. 

\section{Few-fermion system}

As the first example, let us consider a one-dimensional system of well-defined number of fermions of equal mass $m$ belonging to two different components $\sigma\in\{\uparrow,\downarrow\}$ which are confined in a parabolic potential of frequency $\Omega$. We assume that interactions between particles are the simplest possible, {\it i.e.}, they have a contact form between opposite-spin particles. It means that the whole Hamiltonian of the system is a linear function of interaction strength $g$ and it can be written as $\hat{H}(g)=\hat{H}_0+g\hat{H}_I$, where the single-particle Hamiltonian $\hat{H}_0$ and the interaction Hamiltonian $\hat{H}_I$ read
\begin{subequations} \label{Ham}
\begin{align}
\hat{H}_0 &= \int\!dx\,\hat\psi^\dagger_\sigma(x)\left(-\frac{\hbar^2}{2m}\frac{d^2}{dx^2}+\frac{m\Omega^2}{2}x^2\right)\hat\psi_\sigma(x), \\
\hat{H}_I &= \sqrt{\frac{\hbar^3 \Omega}{m}}\int\!dx\,\hat{n}_\uparrow(x)\hat{n}_\downarrow(x).
\end{align}
\end{subequations}
Here, the fermionic operator $\hat\psi_\sigma(x)$ annihilates $\sigma$-component particle at position $x$ and $\hat{n}_\sigma(x)=\hat\psi^\dagger_\sigma(x)\hat\psi_\sigma(x)$ is the $\sigma$-component local density operator. It is clear that in this convention both parts of the Hamiltonian have a dimension of energy, thus interaction strength is controlled by a dimensionless parameter $g$. Although it is not crucial for this discussion, it is worth underlining that physical systems described by the Hamiltonian $\hat{H}(g)$ are attainable in the state-of-the-art experiments with ultra-cold atoms and they can be controlled with tremendous accuracy (for review see \cite{2012BlumeRPP,2016ZinnerEPJ,2019SowinskiRPP}). Therefore, the discussion is not only academic but have also practical consequences. Since the exact solution of the problem with $N>2$ is known only in two trivial limits, {\it i.e.}, $g=0$ and $g\rightarrow \infty$ (see \cite{1960Girardeau,2001Girardeau,2007Girardeau} for details), one tries to harness all possible numerical methods to obtain the spectrum of $\hat{H}(g)$ as precisely as possible. This is especially challenging for strong interactions ($g\gg 1$) since then interactions dominate single-particle excitations. In this range, the decomposition of any many-body eigenstate always contains an essential contribution from Fock states containing highly excited single-particle orbitals. Thus, in practice it is not possible to handle all information needed for a precise description of the state. In the following, we limit ourselves only to the problem of the many-body ground state since this case already exposes the bottleneck of the eigenvalue continuation in all its glory. However, generalization to other many-body eigenstates is straightforward.  

According to the eigenvalue continuation scheme, to get the ground-state wave function and its energy for some target interaction strength $g_\odot$, first we choose $K$ different values of interactions $\{g^{(1)},\ldots,g^{(K)}\}$ for which corresponding many-body ground states $\{|G^{(1)}\rangle,\ldots,|G^{(K)}\rangle\}$ and their energies $\{E^{(1)},\ldots,E^{(K)}\}$ can be determined with appropriately high accuracy. In the case studied, we pick $K=4$ interaction strengths from the vicinity of the non-interacting system, $g\in\{0.1,0.2,0.3,0.4\}$, for which the exact diagonalization of the many-body Hamiltonian can be performed very accurately. The diagonalization is performed in the Fock basis $\{|F_i\rangle\}$ of non-interacting many-body states having the lowest energy, {\it i.e.}, from the infinite set of all many-body eigenstates of the Hamiltonian $\hat{H}_0$ we select only these whose energy is not larger than a fixed energy cut-off $\mathtt C$. Of course, along with increasing cut-off $\mathtt C$ the accuracy of the final result is improved. However, since we consider here quite weak interactions, we find that for large enough $\mathtt C$ further increasing of the cut-off does not change the result noticeably. A detailed prescription of this approach can be found in earlier works \cite{1998HaugsetPRA,2007DeuretzbacherPRA,2019ChrostowskiAPPA,2020RojoMathematics,2022RammelArxiv}. After all, as a result of this procedure we obtain all four ground-states $|G^{(k)}\rangle$ as a set of decomposition coefficients in the Fock basis, $|G^{(k)}\rangle = \sum_i \alpha_i^{(k)} |F_i\rangle$.  
Then, to obtain the many-body ground state and its energy for the target interaction $g_\odot$, we apply standard eigenvalue continuation prescription \cite{2018FramePRL}. First, we calculate all matrix elements of the overlap matrix ${\cal N}_{k,k'} = \langle G^{(k)}|G^{(k')}\rangle=\sum_i \bar\alpha_i^{(k)}\alpha_i^{(k')}$ and the full many-body Hamiltonian at target interaction ${\cal H}^{\odot}_{k,k'} = \langle G^{(k)}|\hat{H}(g_{\odot})|G^{(k')}\rangle$. As a result we obtain two $4\times 4$ matrices which in principle contain maximal information about the many-body ground state at target interaction which is encoded in the sampling states and can be efficiently exploited via the eigenvalue continuation approach. Indeed, after solving a generalized eigenproblem for the Hamiltonian matrix ${\cal H}^{\odot}$ with respect to the overlap matrix ${\cal N}$, $\left({\cal H}^{\odot}-E_0{\cal N}\right)|G_\odot\rangle = 0$, one obtains the target ground state as a simple decomposition in the sampling ground-state basis, $|G_\odot\rangle = \sum_k \gamma_k |G^{(k)}\rangle$, and its corresponding eigenenergy $E_\odot$. 

\begin{figure}
\includegraphics[width=\linewidth]{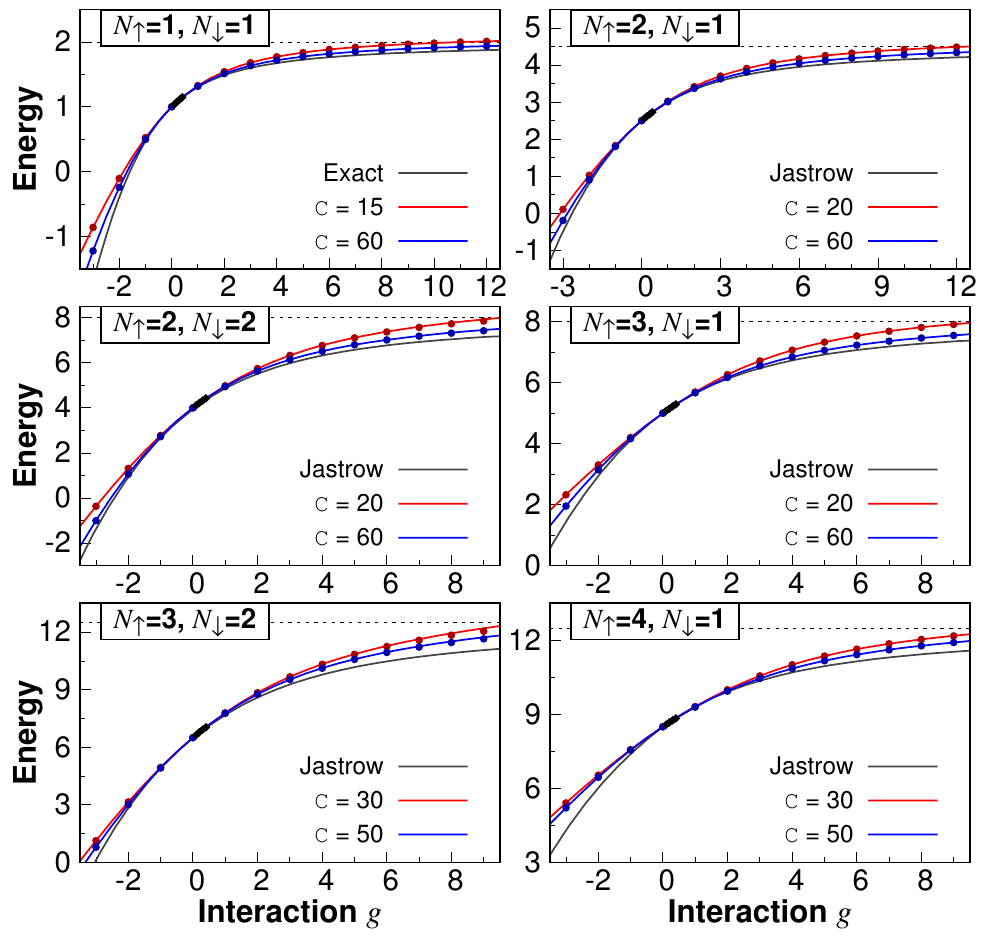}
\caption{Ground-state energy (expressed in natural units of the harmonic oscillator $\hbar\Omega$) as a function of dimensionless interaction strength $g$ for systems with a different number of particles and obtained with different approaches. Black dots represents energies obtained for small interactions $g\in\{0.1,0.2,0.3,0.4\}$ with exact diagonalization method for two different energetic cut-offs of the Fock basis $\mathtt{C}$. Then, these energies are continued according to the eigenvalue continuation prescription (solid red and blue lines). These predictions are compared with results obtained with exact diagonalization (red and blue dots) and a very precise variational scheme based on the Jastrow pair-correlation ansatz. In the case of $N_\uparrow=N_\downarrow=1$ we compare with the exact analytical solution \cite{1998BuschFoundPhys}. Note, that predictions served by eigenvalue continuation never improve direct exact diagonalization results. \label{Fig1}}
\end{figure}

We apply the scheme described above to the system studied containing  different numbers of particles up to five and for different target interactions from a large range. The results for the ground-state energy are presented in Fig.~\ref{Fig1} where we compare predictions of the eigenvalue continuation approach gaining from the exact diagonalization at sampling interactions (solid red and blue lines for different cut-offs $\mathtt{C}$) with the method based on  direct diagonalization of the Hamiltonian of the system. For a clear comparison, the latter is performed on the same Fock basis as the sampling eigenstates for  which the eigenvalue continuation were determined. It is clear that the eigenvalue continuation method works perfectly and is able to extrapolate low-interaction results to very strong repulsions and attractions, far from the perturbative regime. Unfortunately, the results obtained are never more accurate than the results obtained directly by the exact diagonalization method. Moreover, for strong interactions, the method clearly overestimates the ground-state energies of the system with infinite repulsions (horizontal dashed line). In fact, in this particular case, there is no additional gain from using the eigenvalue continuation method and its predictions are fairly worse than results obtained by other, much faster, and more suitable approaches. As an example, with black lines, we present the ground-state energies obtained with the Jastrow-like variational approach \cite{2018KoscikEPL,2021KoscikSciRep}. These comparisons show that increasing accuracy of the sampling states (forced by increasing the cut-off $\mathtt C$), although does not change their energies noticeable, significantly improves the eigenenergy of the target ground state predicted by the eigenvalue continuation method. 

The results clearly signal that the main obstacle and essential limitation for the eigenvalue continuation approach come from its impossibility to go beyond (even approximately) the Hilbert subspace in which the sampling states are calculated. In the case studied, their accuracy is almost perfect since contributions from highly-excited Fock states not included to initial calculations are negligible. However, when interactions are increased, the importance of these neglected states becomes significant and their omission leads to essentially wrong results. Although the ground state is always isolated from other states and smoothly changes with interactions $g$, it always successively flows out of the initial Hilbert subspace; independently of how large and accurate the sampling problem is. This observation should be taken into account when the eigenvalue continuation approach is applied to problems for which structures of many-body eigenstates are not sufficiently known. In the case studied, this effect is directly triggered by the form of mutual interactions assumed. Since contact interactions give rise to cusps in the ground state wave functions at the positions where two interacting particles meet, one needs to extend considered Hilbert space and include additional many-body states to describe the ground state appropriately with increasing interaction $g$. Consequently, the problem is inherently intractable by the eigenvalue continuation. In other words, there is no path leading to the desired target state that can be followed from the sampling points. 

\begin{figure}
\includegraphics[width=\linewidth]{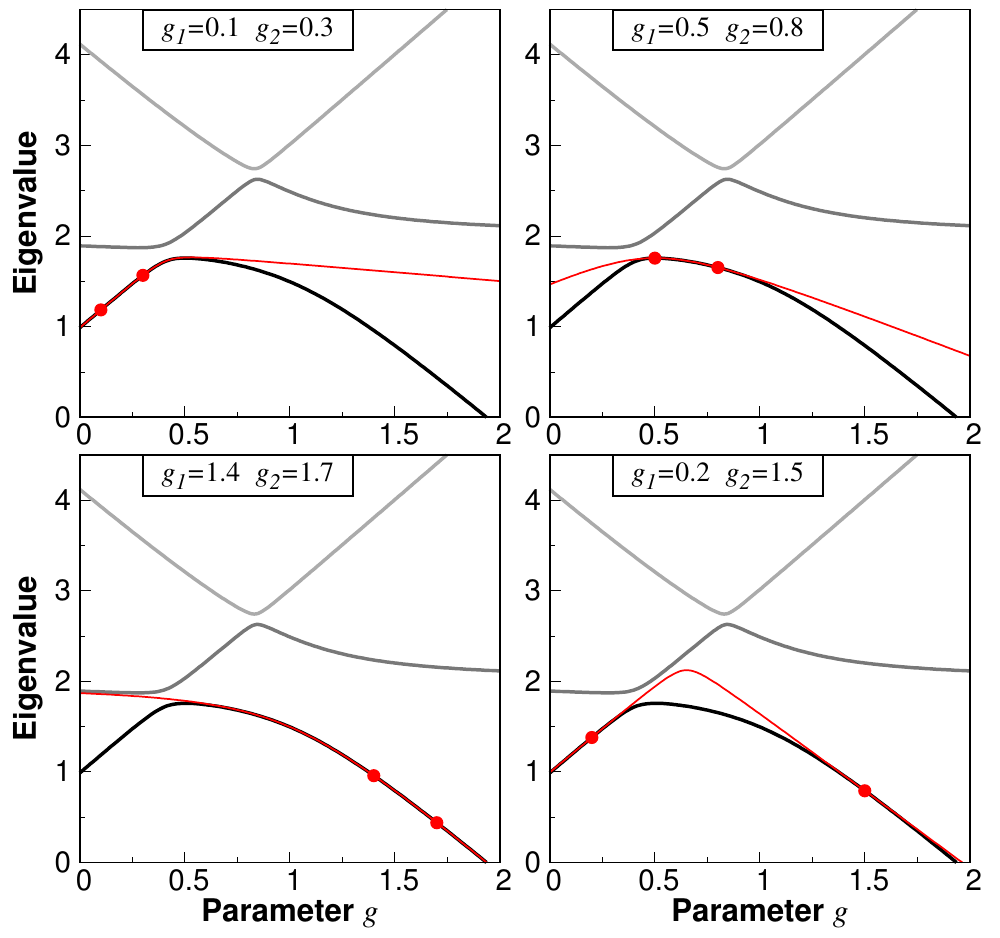}
\caption{ The spectrum of the toy three-level model \eqref{ToyHam} as a function of interaction parameter $g$. In subsequent plots, with red lines, we present the ground-state energy as predicted by the eigenvalue continuation method depending on the choice of two sampling ground states (red dots) obtained exactly for interaction parameters $g_1$ and $g_2$. It is clear that independently of this choice, the eigenvalue continuation cannot reproduce appropriately ground-state energy in the entire range of the coupling since for some interactions $g$ the exact ground state (thin black line) has an essential contribution from the state being perpendicular to the two sampling states. Consequently, the ground-state energy is significantly overestimated by the eigenvalue continuation approach. \label{Fig2}}
\end{figure}

\section{System of three coupled states}

In fact, one can construct different simple models containing all the elements sufficient to expose the main obstacle for the eigenvalue continuation approach. For example, let us consider a simple three-level system described by the Hamiltonian $\hat{H}(g)=\hat{H}_0+g\hat{H}_1$ where Hamiltonians $\hat{H}_0$ and $\hat{H}_1$ are represented by following matrices:
\begin{equation} \label{ToyHam}
H_0 = \left(
\begin{array}{ccc}
1 & 0.1 & 0 \\
0.1 & 2 & 0.5 \\
0 & 0.5 & 4 
\end{array}
\right), \qquad
H_1 = \left(
\begin{array}{ccc}
2 & 0 & 0 \\
0 & 0 & 0 \\
0 & 0 & -2 
\end{array}
\right).
\end{equation}
The exact spectrum of the system can be obtained straightforwardly for any interaction $g$ and it is presented in Fig.~\ref{Fig2} (solid thin lines). We have chosen this Hamiltonian intentionally to illustrate the limitations of the eigenvalue continuation method. As in the original example \eqref{Ham}, the model fulfills all the conditions required for the applicability of the eigenvalue continuation method. In particular, there are no crossings between many-body levels (counterparts of quantum phase transitions in many-body systems) or any breaks of analyticity across the extrapolation. It is clear that all eigenvalues of the Hamiltonian \eqref{ToyHam} are smooth functions of the coupling parameter $g$. Let us also mention that the model, although very simple, is not very far from physical realizations with three-level atomic systems that can be almost perfectly engineered with state-of-the-art quantum optics experiments.

From Fig.~\ref{Fig2} it is evident that, if we perform the eigenvalue continuation in the subspace spanned by two sampling states (marked by red dots), the target state (solid red line) can be well-captured only for some interactions $g$, {\it i.e.}, interactions for which the target state has no essential contribution from the perpendicular subspace. It is clear that in the case studied it is not possible to choose two sampling states in the way that the target state would reproduce the exact ground state in the whole range of interactions. The reason is that along with varying interaction $g$ the ground state always flows out from the subspace of sampling states and this information can not be retrieved by the eigenvalue continuation from anywhere. The only possibility to patch this problem is to include an additional state to the sampling basis. But this will make the eigenvalue continuation method useless, since its complexity becomes equivalent to the complexity of the whole problem of the three-level system. On a much larger scale, the same mechanism of inaccuracy generation is present in the original problem of a two-component mixture of several fermions \eqref{Ham}. Probably, it is also the generic problem for a large class of many-body problems describing interacting quantum particles.


\section{Conclusions}

The eigenvalue continuation method has many advantages and in many cases it serves as an alternative, accurate, and fast method for determining eigenstates of complicated Hamiltonians \cite{2018FramePRL,2021SarkarPRL}. One should remember, however, that its efficiency crucially depends on the quality of the sampling states used for extrapolation to other values of the control parameters. Moreover, it quickly looses its accuracy if the target eigenstates have tendency to flow out from the Hilbert subspace spanned by chosen sampling states.This is not evident in many systems, where the identification of this limitation may be difficult. We illustrate this in a toy model only with three levels, where the third level does not play any role for one set of parameters, but its inclusion becomes crucial in the extrapolated region of parameters. In our first example, for a systems of few interacting fermions, it is more difficult to diagnose this limitation, because as the interactions are increased more eigenvectors has to be included. We show that indeed the accuracy cannot be better than exact diagonalization.

The main obstacle of the eigenvalue continuation method is that, in contrast to other numerical methods, it is not a self-converging method, {\it i.e.}, it gives no tool to affirm that the results obtained are inaccurate or simply wrong. Moreover, even if the problem is detected, it does not serve any prescription for increasing the accuracy of the target state since this would require appropriate capturing of an additional sampling state close to the target one. Without support of other methods, it is not feasible. 

All these limitations of the eigenvalue continuation method are closely related to the orthogonality catastrophe phenomenon \cite{1967Anderson,1967Andersonb} which has been identified for variety of systems of few interacting fermions as well as bosons \cite{2011Goold,2014Campbell,2020Fogarty}.  

In summary, the eigenvalue continuation approach has some fundamental limitations when, along with changes of the control parameter, eigenstates of the Hamiltonian (particularly its ground state) flow into unknown areas of the Hilbert space that are not adequately captured by the sampling states. Related methods as generalized reduced basis methods \cite{2013Lassila,2015quarteroni} also shows this limitations. Unfortunately, this ailment cannot be detected solely within the eigenvalue continuation method and requires support from other, complementary approaches.

\acknowledgments
TS acknowledges fruitful discussions and hospitality at the UPV in Val\`encia. This work was supported by (Polish) National Science Centre Grant No. 2016/22/E/ST2/00555 (TS). MAGM acknowledges funding from the Spanish Ministry of Education and Vocational Training (MEFP) through the Beatriz Galindo program 2018 (BEAGAL18/00203) and Spanish Ministry MINECO (FIDEUA PID2019-106901GBI00/10.13039/501100011033). We thank Pablo Giuliani, Edgard Bonilla, and Kyle Godbey for useful comments on our work and bringing to our attention the connection to generalized reduced basis methods.

\bibliography{biblio}

\end{document}